# An Exploration of User and Bystander Attitudes About Mobile Live-Streaming Video

**Cori Faklaris, Asa Blevins, Matthew O'Haver, Neha Singhal, and Francesco Cafaro**
Department of Human-Centered Computing, School of Informatics and Computing,
Indiana University-Purdue University Indianapolis, Indiana, USA
{cfaklari, ablevi, maohaver, singhneh, fcafaro}@umail.iu.edu

**ABSTRACT**
Thanks to mobile apps such as Periscope and Facebook Live, live-streaming video is having a moment again. It has not been clear, however, to what extent the current ubiquity of smartphones is impacting this technology's acceptance in everyday social situations and how mobile contexts or affordances will affect and be affected by shifts in social norms and policy debates regarding privacy, surveillance and intellectual property. This ethnographic-style research explores familiarity with and attitudes about mobile live-streaming video and related legal and ethical issues among a sample of "Middle America" participants at two typical outdoor social events: sports tailgating and a rooftop party. *In situ* observations of n=110 bystanders to the use of a smartphone, including interviews with n=20, revealed that many are not fully aware of when their image or speech is being live-streamed in a casual context and want stronger notifications of and ability to consent to such broadcasting.

**Author Keywords:** Live streaming; Facebook Live; mobile video; privacy; surveillance; intellectual property.

**ACM Classification Keywords:** K.4.0 [Computers And Society]: General.

**INTRODUCTION**
Live-streaming video is having a moment again. While the ability to deliver video content in real time via the Internet has existed for many years now, the introduction in 2015 of the mobile live-streaming apps Meerkat and Periscope [48], spread rapidly thanks to their integration with the Twitter social networking service and improvements in streaming video technology [24], has sparked renewed growth and user enthusiasm for the medium [29, 31, 34, 38]. Facebook so far is the largest company to jump into mobile live-streaming video [32]. A Cisco white paper predicted that video will account in 2017 for 30 percent of Internet traffic, including 70 percent on mobile devices [36, 49].

Platforms for live-streaming have existed for years, with companies such as Twitch and Ustream broadcasting to a predominantly desktop-based online audience. HCI researchers have also explored privacy perspectives among bystanders to augmented-reality glasses [12] and feeds from stationary always-on cameras used by telecommuters [37].

What is novel and provocative to privacy norms with today's apps, however, is the mobility, personal immediacy and spontaneity afforded by the internet-connected smartphone and the normalization of their use to document the owner's life in almost any context thanks to the popularity of Instagram, Snapchat and the other social networking services in this space. A few prior studies have explored attitudes about the use of mobile live-stream video apps, but they were limited to early adopters [45] and young college students [44].

In this paper, we present the results of an ethnographic study designed to investigate people's familiarity with and attitude towards mobile live-streaming technologies. We first give an overview of notable media and public-policy controversies at the time of our research and of related work in the HCI literature. We then report data from n=110 observations of and n=20 interviews with bystanders among a sample of "Middle America" participants at two typical outdoor social events: tailgating before a sports game and a rooftop party. We discuss key findings and offer design ideas on how to aid the reported concerns of participants.

**BACKGROUND: NOTABLE MEDIA AND POLICY CONTROVERSIES**
Many people now carry in their pockets or purses a portable broadcast studio that professionals once could only dream of. The widespread adoption of internet-connected smartphones that includes a high-quality video camera and software enabling them to almost instantaneously post content online [39], along with the spread of wearable cameras such as the GoPro or Google Glass [4, 5], dashboard cameras [51] and the like, have helped popularize amateur recordings in situations where a traditional stationary video camera would not have been considered practical or acceptable. These mobile and wearable devices already have sparked debates involving individuals' right to privacy [40], to publicity [42] and (in the European Union) to be forgotten [43], as well as their owners' legal rights to film others in various settings, who may view to these videos, and what is captured by the video [2, 26, 27, 40].

As such technologies have proliferated in everyday life, so have the media and policy controversies arising from their misuse and involvement in criminal acts and violent events. The following examples are of particular note when understanding the larger environment in which people are negotiating privacy and other legal and ethical issues with mobile live-streaming video.

**Citizen Recordings of Police Actions**
The wide dissemination of recordings of violent police actions has become a flashpoint for the African-American community. In July 2016, just a day after bystanders video-recorded police officers fatally shooting Alton Sterling in Louisiana, a woman used Facebook Live to stream video in real time of a police officer shooting her boyfriend, Philando Castile, during a seemingly routine traffic stop in Minnesota [35]. "I wanted it to go viral so the people could see," said the woman, Diamond Reynolds. "I wanted everybody in the world to see what the police do." [35] Reynolds' initial broadcast was viewed more than 1 million times before Facebook first pulled it, then reposted it with a warning about graphic content [35].

In most such encounters, video recording or live-streaming is legally permissible under U.S. law because neither police nor citizens have a reasonable expectation of privacy. Mickey Osterreicher, general counsel of the National Press Photographers Association, observed that "citizens can record police and police can record citizens when either is out on the street in a public place" [15]. However, an arrest for invasion of privacy or trespassing may be justified when the setting for video recording or live-streaming is semipublic or private.

**Upskirt Videos, Revenge Porn and Nanny Cams**
Another notable public concern to consider with live-streaming video is that of voyeurism, defined as the "act of filming or disseminating images of a person's *private areas* under circumstance in which the person had a reasonable expectation of privacy regardless of whether the person is in a private or public area" [*see* 2, p. 8]. Notable examples that have been facilitated by the growth in mobile video-recording capability include "upskirt" and "downblouse" videos [3] and "revenge porn" [8]. Devices such as "nanny cams," laptop cameras, modern closed-circuit video systems and other small, unobtrusive wireless recorders have given people a broad ability to put others under surveillance without their explicit knowledge or consent [11].

While recordings such as of nannies or other types of employees, rental occupants and guests via hidden cameras in homes or on other private or semi-public property is generally allowed under U.S. federal and state laws, some caveats apply: first, that the recording be video-only, as those involving audio generally violate federal wiretapping laws [9]; and second, that such recordings be limited to common areas and not bedrooms or bathrooms, where a more absolute expectation of privacy holds [9]. In some cases, such surveillance devices may be welcomed if and when its subjects become explicitly aware of them depending on whether they seem to benefit from their existence. Examples include those that allow video conferencing or are part of security systems for home or business [11, 40].

Video voyeurism is increasingly becoming accepted as legal, if controversial, entertainment as well as big business. Websites such as Chatroulette and YouNow allow users of internet-connected cameras to live-stream video to random strangers around the globe of even mundane activities such as eating and sleeping. Adult video sites such as YouPorn and PornHub will likely continue as the largest internet category of entertainment and business [46]. Two other categories of multimillion-dollar entertainment businesses for live video streaming are gaming and sports [14].

**Live-Streaming of Professional Sports**
The increasing capability to live-stream video for individual smartphone users, as opposed to broadcast and cable television networks, poses a threat to revenue streams for live events such as professional sports [14, 36]. During the 2015 boxing match between Floyd Merryweather and Manny Pacquiao, it was estimated that more than 10,000 people viewed a Periscope stream of the Las Vegas bout instead of paying for Showtime, HBO or a video-on-demand stream of the fight [30]. The overall revenue from paying audiences was more than $400 million, raising questions of how much more could have been earned if these app viewers had not been able to access it for free [30].

But mobile live-streaming video is also providing commercial sports enterprises a way to directly broadcast these events, bypassing traditional television intermediaries, and to engage with fans on social media [30]. Indeed, Periscope's parent, Twitter, had already moved to strike deals for broadcasting professional sports at the time of our research [28]. The rapid rollout of Facebook Live among that company's media partners in 2016 further normalized the direct video streaming of live events via social media.

**RELATED WORK**
Privacy considerations with video streaming are not new: they have been previously explored in the context of wearable cameras, telecommute, and pre-recorded video sharing. Understanding the privacy implications of video-streaming technologies is, however, still a challenge.

**Wearable Cameras**
Denning et al. [12] have conducted ethnographic-style research into the privacy perspectives of individuals who were bystanders in the vicinity of augmented reality devices modeled on Google Glass. During 12 field sessions in local cafes, they observed and interviewed 31 such persons regarding a mock AR device that was worn publicly by the researchers. Participants noted that what was being recorded made a difference in their perceptions [12], which echoes the discussion in Bohn et al. [7] of the actor and the context as two components of perceived boundary violations in ubiquitous computing. Participants were interested in being asked for permission to record and the ability to block transmission [12]. Denning et al. used the results to sketch potential design axes for privacy-mediating technologies, such as push/pull, opt-in/opt-out, place-based versus proximity- or identity-based, and user versus bystander or third party [12]. This is congruent with Erickson and Kellogg's work in identifying the factors of awareness, visibility and accountability in designing for social translucence [16]. These recommendations, however, are specific for wearable cameras, and difficult to generalize to live mobile-streaming. Furthermore, they generally assume that everybody is aware of what wearable cameras are, how they look like, and when they are in-use.

**Cameras and Telecommute**
The difficulty of the design work in balancing privacy with such awareness was shown by Neustaedter et al. in 2006 [37]. In a laboratory experiment, they recorded the attitudes of 20 participants who were shown blur-masked images such as of nude co-workers that might accidentally be transmitted by always-on video feeds of telecommuters working in home environments. The blur-filtering technology was not enough to answer some participants' concerns about "high-risk" privacy violations in a home-to-office video system. The authors suggested additional elements be incorporated for privacy regulation and feedback, such as gesture-activated blocking in proximity to the camera, audio feedback such as the sound of a camera clicking or rotating and visual feedback such as LED lights.

Neustaeder et al. noted the limitation of their privacy findings in the simplified artificial context of their laboratory experiment versus the more-complex contexts in which such concerns are situated in the real world. The home media space also is a context that is a private enclosure and which those present in the home make a choice to enter despite knowledge of the presence and the risks of the always-on video feed.

**Pre-Recorded Mobile Video Sharing and the Social Web**
The possibilities for mobile-broadcast live video shared on the social web, as distinct from mobile video calling, website-based streaming and desktop-original platforms such as YouTube, were explored in 2010 by Juhlin et al. [29]. In a qualitative analysis of n=178 video clips posted to four websites, they sorted the clips into the following topics by number of occurrences: test broadcasts, screens, groups and crowds, tours, social events, kids and pets, demonstrations, presentations, performances, video logs, landscapes and sudden events (as well as a "not viewable" category of video misfires) [29]. The authors focused on shortcomings in the novice videographers' techniques and in the lack of camera interface cues such as a countdown timer from the time the record button is pressed to the broadcast of the first image frame, suggesting that these would need to be addressed in order to fulfill the medium's promise for empowering citizens and democratizing video broadcasting. Their work did not address whether the seeming shortcomings were a motivating or limiting factor in their broadcast and consumption, nor did it address any legal or ethical issues such as invasion of privacy with domestic videos or intellectual property concerns with the broadcast of artistic performances or tours.

The following year, Dougherty [13] offered an analysis of n=1,000 randomly selected mobile videos posted over five months on Qik.com and interviews with n=7 producers of these videos. She noted "spontaneity" and "immediacy" as motivating factors for mobile live video sharing. Those interviewed for her study also cited a motivation of building an audience for more civically minded videos such as of school board meetings through their sharing of personal or otherwise unsophisticated video clips. Her interviewees reported being mindful of privacy issues with public filming and of general ethical issues in video production. Somewhat contradictorily, her content analysis revealed many videos shot by men were of women in domestic spaces or of coworkers of either gender and that some of these subjects may not have been fully or even partially aware of being captured on video for an audience. Dougherty did not explore the familiarity or attitudes of bystanders to such video broadcasting.

**Recent Inquiries Into Mobile Live-Streaming Video Apps**
Tang et al. [45] appear to be the first HCI researchers to investigate the use of the current generation of dedicated apps for mobile live-streaming video. In their interviews with n=20 early adopters of Meerkat and Periscope and crowdsourced analysis of n=767 live streams on these apps during April-May 2015, they found that a significant amount of use can be characterized as either personal blogging or branding. Coders on Amazon Mechanical Turk identified many videos as featuring *expository* content (chatting) as well as *experiential* content (broadcasts of notable objects, places or events). Users said they relished the immediacy of the connection with their audiences, who often interact via emoji and text comments with the live streamer during the broadcast. While many users said they found the apps to be easy to learn and use, they described investing considerable time and thought into deciding what to broadcast, how to present themselves on camera and in the app, how to cultivate followers and how best to interact with those watching and deal with inappropriate comments – community work that is common among all forms of SNSs.

The Tang study, however, did not include any research that involved bystanders to mobile live-streaming video. The authors also noted the need for continuing research as the apps matured and evolved in the marketplace.

In 2016, Singhal et al. [44] presented findings from an *in situ* field study of the reactions of bystanders in an indoor university building to two types of devices for capturing video: a smartphone and Google Glass. Their interviews with n=9 of these bystanders revealed that many expressed fewer concerns with streaming video than with recorded video because of its perceived ephemeral quality; participants thought that streamed video was not likely to be saved to a disk and that their activities, even if embarrassing in nature, would be visible online for a few seconds only. They were also more likely to accept video capture from a smartphone in a public place and if the camera was constantly moving rather than fixating on them. Additionally, three of the four female participants said they would be more uneasy if a male was using the camera. All participants said they wanted to be asked for permission before video was recorded of them.

The generalizability of Singhal's study, however, is limited because of the small sample of a relatively homogeneous population (all were students, all were 25 or under), the field studies' location in the relative artificiality of an indoor campus environment, and the researchers' choice to hold the smartphone in a horizontal orientation for their study. It may seem like a small point, but this orientation is an anachronism now that Snapchat has popularized vertical video capture on smartphones in casual contexts: this makes it more difficult to identify potential streamers among people who have a mobile phone in their hands.

## PROBLEM SPACE

As the above review notes, the popularity and pervasiveness of mobile video has brought with them numerous legal and ethical concerns for their use. The existing literature, however, either focuses on the design of technology itself or, when the focus is on users and passersby, seems to abstract from the complexity of everyday live mobile-streaming scenarios. Specifically, we believe that an investigation on attitudes and behaviors of the people who are interacting with live mobile-streaming application needs to: (1) assess people's familiarity with live mobile-streaming technologies and with the legal implications of their use; and, (2) consider social space (public, or semi-public), device, app, user(s), and bystanders, as a unit of analysis.

Our research was designed to remedy this gap with data collected for the following research questions:
- *RQ1.* What familiarity and legal or ethical attitudes do likely users and subjects/ bystanders report regarding mobile live-video technologies and other apps or platforms?
- *RQ2.* How do bystanders react "in the wild" to the presence of mobile live-streaming video in public and semi-public settings?

## METHODOLOGY

For our research into the above questions, we planned and executed qualitative field studies to gauge how awareness of and attitudes about mobile live-streaming video varied among bystanders to simulated broadcasts "in the wild."

### Participants

All participants for the field studies were encountered or recruited in August 2016 in a U.S. Midwestern metropolitan area of more than 2 million, and was diverse as to age, race and education. Our participants were people who are current U.S. residents and age 18 or older (although video apps are popular with children and teens, we believe that issues with mobile streaming and informed consent from minors may require an ad-hoc study).

In contrast with Tang et al. [45], our study largely is not comprised of expert users or "early adopters" of the technology being studied – we did not want to assume that both users and passersby are always completely aware of live mobile-streaming applications. Only 4 in 20 (20.0%) of those we interviewed told us they had ever live-streamed video. However, our participants were not strangers to computing devices in general, as almost every interviewee reported at more than 40 hours a week of such use overall, whether desktop, laptop, tablet, smartphone or wearable device.

Half of the interviewees ($n_1$=10) were encountered randomly during tailgating prior to professional soccer games, while the other ($n_2$=10) participants were recruited by a co-investigator for what they were told was a focus group, convened on the outdoor party deck of an urban condominium building. The exact demographic breakdown of these participants in total is as follows: Male n=13, female n=7; age 18-29 n=8, age 30-39 n=6, age 40-49 n=4, age 50-59 n=2; graduate degree n=11, bachelor's degree n=4, some college, technical degree or associate's degree n=5; Caucasian/European heritage n=14, Asian n=3, African n=2, Latino n=1.

**Procedure**

We conducted field studies to measure how a total of n=110 bystanders reacted "in the wild" to the presence of mobile live-streaming video in the following social/spatial contexts: an outdoor gathering space that is open to the public (Phase 1); and, a semi-private, controlled-access meeting space (Phase 2). Of these, we recorded a total of n=20 interviews.

*Phase 1: Public Gathering Space*

On two separate occasions in August 2016, we conducted field observations and interviews of tailgating fans gathered in a public outdoor space before the start of home games for a local professional soccer team. Specifically, these inquiries were located in the open-air surface parking lot and along the tree-lined sidewalks outside the "will-call" box office of the stadium. We posted notices and study information sheets on posts and trees around the perimeter of the area at least two hours before game time to let passers-by and entrants to the space know that they were being video-recorded as part of a research study, though without specifying what that study was. Other parking areas and sidewalks nearby were available for any tailgating fans or other passers-by that did not wish to enter the space and be subject to our video recording.

We chose to simulate the act of streaming live video from a mobile phone, rather than actually broadcast live, in order to protect against the type of privacy violations that are a subject of our study. Accordingly, a team member started walking around with a mobile phone, pantomiming the act of recording live-streaming video by holding her phone up in a vertical orientation and narrating the scene as if she were broadcasting to an unseen audience. Meanwhile, a second team member stood a distance away with a portable video camera to record bystanders' reactions to the simulated mobile video. From analyzing the video footage, we estimate that a total of $n_1$=90 bystanders were recorded in proximity to these pantomimes.

After several minutes of this pantomime around the space, two other team members approached and invited random bystanders to participate in semi-structured interviews about the "live-streaming," using the following questions adapted from Denning et al. [12]:
- *Did you notice the person who was broadcasting live video from their phone? What about them did you notice?*
- *Have you heard about mobile live-streaming video apps like theirs? What have you heard?*
- *Why do you think someone would want to stream live video from their phones?*
- *How do you feel about being around someone who is streaming live video? Why?*
- *Would you want someone to ask your permission before streaming a live video from your location?*
- *Would you be willing to take an action to block someone from being able to stream video of you?*

During these interviews, our video camera operator recorded the conversations, while the interviewers recorded audio of the participants and took written notes on their demographics and the answers to the above questions. Interviews were later transcribed. A total of $n_1$=10 bystanders were interviewed for an average of 7 minutes each on the two occasions of these *in situ* studies. Each interviewee was given a $5 gift card as compensation for their time.

*Phase 2: Semi-Private Meeting Space*

In our last field study, also in August 2016, we recruited $n_2$=10 participants to convene with four co-investigators and a confederate at a semi-private meeting space on private property (the rooftop deck area of a condominium building). Access to this building is restricted to residents and their guests, though people inside the building are free to enter the deck area at any time. Our team secured the permission of the management for this use.

On the day of our study, a team member posted a notice and study information sheet outside the door to the deck area to alert anyone entering that they would be subject to video-recording as part of a research study, though as with Phase 1, without giving study specifics. Likewise, our recruits were informed that they would be recorded during a focus group and interviewed about the social implications of technology, but were not provided with the specific apps or research questions under investigation.

One co-investigator was stationed in the corner with a video camera to record the interactions among participants. Each invited participant was greeted by three other co-investigators and provided with seating and refreshments (valued at an estimated $5 per participant). Our confederate also was greeted and introduced as an 11$^{th}$ participant in the proceedings. Once the group was assembled, we introduced a misdirection by setting up a portable white board, asking participants to pull out their smartphones and then leading them in a discussion of why they favored certain mobile apps for keeping connected with others.

While this activity was being conducted, our confederate, who had been sitting with the participants and took out her smartphone along with them, began simulating the act of live-streaming video using her phone, first from a seated position and then while walking around the perimeter of the study space. As with the previous field studies, the confederate held her phone in a vertical position and narrated the scene to an unseen audience. Once she had made two circuits, we stopped, introduced her fully and informed the gathering of the exact purpose of the study.

The co-investigators then interviewed the participants individually, using the same questions adapted from

Denning et al. [12] in Phase 1. The interviewers recorded audio of the participants and took written notes on their demographics and the answers to the above questions. Interviews were later transcribed.

**Data Analysis**
After each field session, three researchers conferred to debrief each other and to write down impressions while they were fresh in mind. Once the research was concluded, we collected together the video recordings of each field session that documented reactions to the pantomimed live-streaming, the code sheets with written demographic information about each interviewee, the audio recordings of these interviews and the supplemental written notes from the field studies. To analyze the bystander observations, one researcher watched the video recordings of the field sessions to count how many people were in the vicinity of the pantomimed smartphone streaming and to count how many of those nearby reacted in a visible physical fashion to the pantomiming confederate, either with glances or by turning their heads or bodies toward the action or overheard conversation.

For the interviews, the audio recordings were transcribed and read through to identify themes and commonalities, which were discussed and agreed to by the entire research group.

**RESULTS**
**Bystander Reactions and their *Awareness* of Live Mobile-Streaming**
Observations at the scenes and analyses of video recorded in Phase 1, the open public tailgating space, shows that not many bystanders ($n_1$=15) visibly noticed the actions of our confederates in their vicinity as indicated through glances, turning their heads or bodies toward the action or overheard conversations. In Phase 2, the semi-private rooftop deck space, all participants ($n_2$=10) reacted with a glance, head or body turn to the pantomime. We theorize this difference is because of the tighter space and more intimate feel of the smaller social gathering in Phase 1 versus the raucous party atmosphere in Phase 2.

Interviews with selected bystanders in both phases confirmed that not many suspected the confederates were live-streaming using an app such as Periscope. Most ($n_1$=6 in Phase 1, $n_2$=6 in Phase 2) thought they were recording video, taking audio notes or using FaceTime or Skype. Some assumed the confederates were playing the just-released mobile game Pokémon Go. Sample comments from interviews:

> [P2] *"We were trying to figure out whether she was recording video or searching for Pokémon."*

> [P4] *"At one point I thought she was talking to somebody."*

> [P20] *"Didn't realize what she was doing. Just thought she was videotaping for later. You can't tell the difference, right?"*

***Familiarity* with the Apps**
About half (n=11, 55%) of interviewees across Phase 1 and Phase 2 said they had heard of mobile live-streaming video apps, and many of these said they had watched at least one video stream on Facebook and seen multiple notifications for them. A general degree of indirect familiarity with such applications transpires from comments that participants made regarding the streaming of public sport events and police actions. For example, interviewees described *experiential* live streams [41], such as those being broadcast from a novel or newsworthy event such as the Rio Olympics, as a positive use of the apps or even a social good.

> [P16] *"If there's news going on right as the person is live-streaming, it's very helpful to see what's going on in the moment."*

This sentiment held even for shocking or violent news events such as the Philando Castile shooting, which was mentioned specifically by three interviewees.

> [P12] *"We need more cops to live-stream. We need more live streaming of the scary things going on."*

> [P9] *"I think [it's good] to get the true content or ideas of things that are really happening live, so there's no way you can deny it. For the simple fact is, it is live."*

Far fewer (n=3, 15%), however, reported being directly familiar with them, they had used one of the apps to themselves live-stream video, with Participant 7 saying "*too much data*" would be used on his phone if he tried it.

**Social Attitudes About Live Streaming**
Surprisingly, most bystanders expressed a general level of comfort with the fact that live-streaming was occurring nearby. Some interviewees (n=8) expressed discomfort with aspects of the confederate's live-streaming pantomime that seemed obtrusive or annoying, such as narration or standing in close proximity to one or two people only. There was no difference in these attitudes across Phase 1 and Phase 2.

> [P14] *"Live streaming for social gatherings [is appropriate], but not when I am walking around [the local entertainment district] by myself.*"

> [P11] *"If they just want to film it, that wouldn't bother me, but if they're narrating while other people are talking, it's [distracting].*

Others (n=5) noted that the act of live streaming is no longer novel and, thus, is easier to ignore.

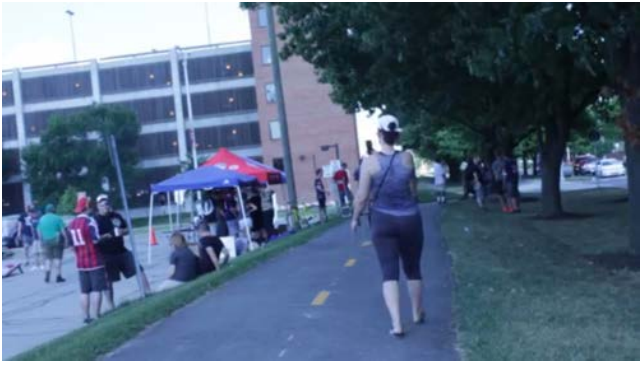

**Figure 1: Pseudo-Periscoper at a Phase 1 field study site.**

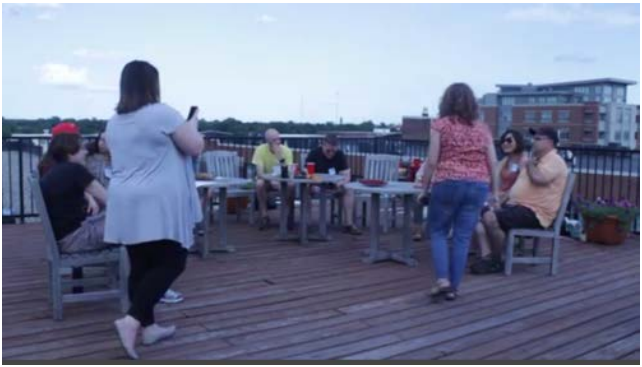

**Figure 2: The scene of the Phase 2 field study.**

[P16] *"I am around it enough now so that it is fairly commonplace. So I am like, whatever."*

[P12] *"I'm desensitized to it. My friends are always on Snapchat. ... We're videotaped all the time now."*

### Attitudes About Appropriateness and Legality

In Phase 1, the public gathering place, interviewees supported live-streaming as an appropriate and legal activity.

[P2] *"Personally I don't really care, I know some people freak out but if you're in a public spot, there's no way to stop them."*

[P6] *"Everybody has a right to privacy, but when you're in a public place, you're in a public place. C'mon."*

In Phase 2, the semi-public meeting space on private property, most interviewees (n=8) were not concerned about their video being recorded and live-streamed; rather, someone expresses reservations on whether using a mobile phone is a good behavior in a professional setting.

[P18] *"I wouldn't feel uncomfortable because I don't think it's that different from taking pictures. It might make me less willing to interact with that person because his or her attention is somewhere else."*

Even those interviewees who had reservations about being on camera (n=3) still saw live-streaming as on par with traditional recording devices in terms of its appropriateness to use in the space.

[P11] *"I tell friends I don't want to be in pictures. I'd also want to know ahead of time if they're live-streaming."*

Interesting, one person was concerned about limited portions of her image or voice being streamed without permission in a different context –in a way that would manipulate or misrepresent her original ideas.

[P17] *"The face thing doesn't bother me so much as if I were talking and that gets posted. I'm very direct and I have opinions. ... Maybe I intended to make that comment in a social setting for 10-15 people, but all of a sudden it's taken out of context."*

### Notification and Permission

Nearly all interviewees indicated that, either verbally or through some sort of technology, they would like to receive stronger notifications of nearby live streaming. They also wanted the ability to consent to taking part. This echoes findings of Denning et al.'s investigation [12] of augmented reality devices and Singhal et al.'s [37] study with students.

[P20] *"Kind of like with Google Glass ... people's concern was, if I'm talking to you and you're videotaping this, I'd like to know that because it could affect what I would say and how I would say it. There are different zones [of behavior]."*

One participant described how it should be "common manner" to ask for permission to people near the camera.

[P15] *"If I were just in the background, I probably won't care. But if I'm the focus of the live stream, I'd like to know what's going on. I would want them to get my permission to do that. ... It's common manners."*

Few interviewee (n=5, 25%) said they would put forth effort such as signing up for a "Do Not Record" registry or taking other initiatives to block or restrict live-streaming of their persons by app users nearby.

[P20] *"I might, if you can turn it on and off."*

[P5] *"It would be helpful ... if you have an option to get out of that stream or to participate."*

# DISCUSSION

## Familiarity vs. Awareness

Although most of our participants (55%) were indirectly or directly familiar with mobile live-streaming apps, a much smaller number of them (less than 25%) actively noticed the simulated use of such technologies through our study. In particular, people's familiarity with a variety of mobile applications (that are not related with live-video streaming) seems to completely normalize the confederate's usage pattern (holding a phone up while narrating a scene) –thus, reducing their awareness on being recorded and their video being (supposedly) broadcasted live. Future work should further investigate this phenomenon.

## Social Norms, Proximity to the Streaming Device, and Permission

One overwhelming, if intuitive, key theme from our work was that bystander interactions with mobile live-streaming video pose challenges that are more social than technical in nature.

For example, P15 describes his request to ask for consent to those who are in close proximity and likely to be featured in the foreground as "*common manner.*" The same interviewee commented regarding appropriate contexts for live-streaming that certain situations such as "*a funeral, an [Alcoholics Anonymous] meeting are both places where live streaming is probably not appropriate and where you don't want to be live-streamed.*" A non-technological solution would be a public awareness campaign to promote the desirability of app users asking for permission and otherwise being sensitive to others in public situations before broadcasting their actions.

We believe, however, that such social norms may be also supported and/or enforced through the design of mobile applications. For example, P15 suggested a smartphone setting that toggled "I don't want to be live-streamed" and "I don't care" to broadcast your preferences to those around you. Furthermore, the idea of spatial boundaries (people in the background vs. in the foreground) echoes the "*zones of interaction*" described by Hall in [23] and the application of proxemics to ambient displays (*e.g.*, [1]) and personal interactive devices [47]. Future work should investigate which spatial boundaries are most relevant for mobile live-video streaming, and the best strategies to enforce them.

## Impact of the Social Space: The Presence of Others as a Normalizing Factor

Interviewees repeatedly drew parallels among their threshholds for boundary violations [7] between *group shots* and *public contexts* versus *tight shots* and *private contexts*. Being streamed as one in a large crowd, usually in a public space and where you are less likely to be identifiable, was seen as much less concerning than being streamed as the sole or among only a couple people in a more intimate or private setting, where disclosure of your presence is almost assured and may be unwanted.

## (Un)Consciousness of the Legal Implications of Mobile Live-Video Streaming and of Copyright Laws

Possible legal consequences to live-streaming were also mentioned by a few interviewees (n=2). Participant 13 said she thought broadcasting from a street protest might be "*a little dangerous*" and that some "*wouldn't want other people to know*" it was their action, taking it down after the fact. In some sense, these comments express disconnect between what is perceived as legal (video-streaming in public spaces) and what is, instead, desirable.

Another area of concern revolves around live events and intellectual property rights for the performances or even for the rights of publicity for the attendees. The question of what is fair use for broadcasting from events, or even for using the likeness of people who can be identified through facial recognition, was mentioned by some interviewees. Participants, however, generally had a positive attitude towards mobile streaming of public events (e.g., P16 described it as a positive use of the apps or even a social good). In addition to clarification in case law and in legislation about the proper use of live streaming at public events involving artistic performances or sports matches, a need exists for more education of the general public about what types of use of new technologies will be allowed under "fair use" exceptions to copyright law.

# DESIGN RECOMMENDATIONS

Although the majority of participants had a generally positive attitude towards mobile live-streaming technologies, some expressed reservations that we believe should be addressed when designing for such technologies. This is not a comprehensive list; rather, our aim is to start a conversation about the issues that participants identified during our study. Specifically, mobile devices should inform those in visual proximity to the devices that they are being recorded (to increase awareness), help all bystanders control the ability of mobile app users to record and broadcast video of them (to allow them to provide permission), and allow to disable mobile streaming remotely when needed (to enforce copyright at public events).

### *Awareness: Colored Lights to Indicate Front or Back Live Video Capture*

Most participants (n=18) said they could not tell that our confederates were live streaming vs. recording notes, snapping photos or playing games. We propose a color LED light next to the camera lens in mobile devices, akin to the red LED light of analog videos cameras that would turn on when the camera was in use, to alert a user or bystander to its activation by lighting up and perhaps blinking while video is being recorded or streamed live over the Internet (Figure 3).

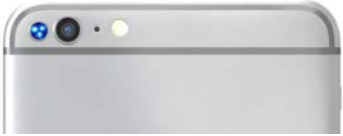

**Figure 3: Proposed back recording light on mobile phones.**

This solution would help people set personal boundaries regarding others' live-streaming, consistent with the concepts of what constitutes violations of these transitory borders as described by Bohn et al. [7] and with design of privacy-mediating technologies that are proximity-based, as originated in Denning et al. [12]. It would supplement verbal notifications of live-streaming and ease concerns about surreptitious surveillance or streaming at live events.

*Permission: "Do Not Record" List and/or Toggle Button*

Participants said that unexpected live streaming by others discomfits them when they are in the foreground or otherwise in close proximity. To provide a low-user-effort method of combating this issue, we suggest a "Do Not Record" database to which individuals can register an image of their face. Mobile streaming apps would be required to check faces in the camera's field of view against the database. If the facial recognition system found a match, the app would blur out the registered user's face. Alternatively, mobile streaming apps might be forced to communicate with nearby devices to check if anybody within the camera field of view has activated a "I don't want to be live-streamed" toggle button on her/his phone.

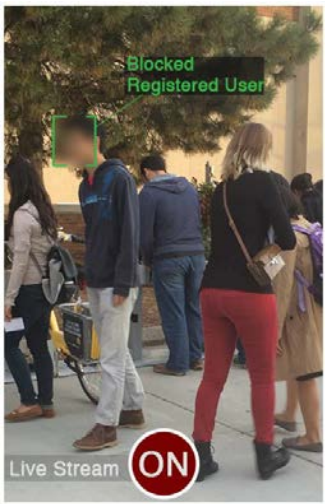

**Figure 4: Passerby face is blurred thanks to non-recording database or toggle setting**

Such a system would be consistent with Bohn et al.'s discussion of boundary management [7] and Goffman's concept of impression management [22], as well as the E.U. legal doctrine of the "right to be forgotten" [43]. Nevertheless, a facial recognition system itself would invoke a host of legal and ethical issues.

*Copyright Laws: Remote Deactivation of Streaming and/or Camera*

Similarly, a functionality to remotely de-activate streaming capabilities may be needed in order to comply with copyright regulations (e.g., during sport events). In the absence of any strong advocates for legal safeguards on permissive public live streaming, it is likely that corporations will set de-facto public policy through technology rollouts. Indeed, Apple has already filed for a patent for an infrared system that could disable iPhone cameras at public events, though it is unclear if they plan to deploy the technology [6].

**CONCLUSIONS AND FUTURE WORK**

In this paper, presented the results of two field studies pertaining to the public's awareness of and familiarity with mobile live-streaming video apps. Our study centered on understanding the emotional response to being live streamed "in the wild," in contexts where people could be recorded without giving permission. We were also interested in gauging the public's interest in protecting themselves from being live streamed and their response to ideas for doing so.

Many interviewees were initially unaware that someone was live streaming in their vicinity, perhaps even broadcasting their image. While our pseudo-Periscoper was trying to be obvious in her exploits, many respondents noted that they could tell she was recording video, talking to someone on Skype, or playing Pokémon Go, but not that she was broadcasting to the world. After being informed of the live-streaming, many participants reported a neutral response—they weren't bothered by her presence or her recording, but they might appreciate the live-streamer informing them if they are going to be on the stream for an appreciable amount of time. We offered design recommendations and hope to inspire more work in this area.

We foresee several directions in which to take this research. We may pursue a wider study of user familiarity with and attitudes about privacy, data and content sharing on various social sites and devices that targets a larger and more diverse population sample. We also may implement our design recommendations for a prototype tool, intervention or affordance that addresses the main points from our work to evaluate whether our ideas would be effective among likely users and bystanders to these technologies.